\documentclass{revtex4}
\pdfoutput=1
 \usepackage{bm}
 \usepackage{graphicx,color}
 \usepackage{latexsym}

\newcommand{\citea}[1]{[\cite{#1}]}
\newcommand{\Pl}{\mathcal{P}}
\newcommand{\I}{\mathcal{I}}

\newcommand{\V}{\mathcal{V}}

\newcommand{\Bk}{\mathbf{k}}
\newcommand{\W}{\mathcal{W}}
\newcommand{\id}{\phantom{i} d}
\newcommand{\J}{\textrm{J}}

\newcommand{\grad}{\mbox{\boldmath\(\nabla\)}}
\newcommand{\dive}{\mbox{\boldmath\(\nabla\cdot\)}}
\newcommand{\curl}{{\mbox{\boldmath\(\nabla\times\)} }}
\newcommand{\dotv}{\mbox{\boldmath\(\cdot\)}}
\newcommand{\cross}{\mbox{\boldmath\(\times\)} }

\newcommand{\Ba}{\mathbf{a}}
\newcommand{\BB}{\mathbf{B}}

\newcommand{\Bj}{\mathbf{j}}
\newcommand{\BJ}{\mathbf{J}}

\newcommand{\BA}{\mathbf{A}}
\newcommand{\Bn}{\mathbf{n}}

\newcommand{\Bb}{\mathbf{b}}

\newcommand{\BX}{\mathbf{X}}

\newcommand{\norm}{M}
\newcommand{\nsing}{\eta}
\newcommand{\Br}{\mathbf{r}}
\newcommand{\Bxi}{\bm{\xi}}
\newcommand{\Brho}{\bm{\rho}}

\newcommand{\lbrac}{\left[\!\left[}
\newcommand{\rbrac}{\right]\!\right]}
\newcommand{\xir}{\xi_r}
\newcommand{\xith}{\xi_{\theta}}
\newcommand{\xiz}{\xi_z}
\newcommand{\Bth}{B_{\theta}}
\newcommand{\Bz}{B_z}

\newcommand{\ibar} {\mbox{$\,\iota\!\!$-}}
\newcommand{\der}[1]{_{#1}^{\phantom{#1}\prime}}

\newcommand{\rhat}{\mathbf{\hat{r}}}
\newcommand{\thhat}{\bm{\hat{\theta}}}
\newcommand{\zhat}{\mathbf{\hat{z}}}

 \newcommand{\eqn}[1]{Eq.~(\ref{eq:#1})}
 \newcommand{\Sec}[1]{Sec.~\ref{sec:#1}}

 \newcommand{\be}{\begin{eqnarray}}
 \newcommand{\ee}{\end{eqnarray}}

 \newcommand{\insertfigure}[3]{
\begin{figure}\begin{center}\includegraphics[width=.45\textwidth]{#1}\caption{#2\label{fig:#3}}\end{center}\end{figure}
 }

\begin{document}

\title{%Resolving MHD Equilibria and Stability of Stepped Pressure Profile Plasmas
Magnetohydrodynamic Stability of Plasmas with Ideal and Relaxed Regions}
\author{R. L. Mills, M. J. Hole, R. L. Dewar}
%\email{ruth.mills@anu.edu.au}
\affiliation{Department of Theoretical Physics and Plasma Research
Laboratory,\\
Research School of Physical Sciences and
Engineering, The Australian National University, ACT 0200 Australia}

\date{\today}

%\pacs{52.30.Cv,52.55.-s}
\begin{abstract}

A unified energy principle approach is presented for analysing the magnetohydrodynamic (MHD) stability of plasmas consisting of multiple ideal and relaxed regions. By choosing an appropriate gauge, we show that the plasma displacement satisfies the same Euler-Lagrange equation in ideal and relaxed regions, except in the neighbourhood of magnetic surfaces. The difference at singular surfaces is analysed in cylindrical geometry:  in ideal MHD only Newcomb's [W.~A. Newcomb (2006) \emph{Ann. Phys.}, \textbf{10}, 232] small solutions are allowed, whereas in relaxed MHD only the odd-parity large solution and even-parity small solution are allowed. A procedure for constructing global multi-region solutions in cylindrical geometry is presented. Focussing on the limit where the two interfaces approach each other arbitrarily closely, it is shown that the singular-limit problem encountered previously [M.~J. Hole \emph{et al.} (2006) \emph{J. Plasma Phys.}, \textbf{77}, 1167] in multi-region relaxed MHD is stabilised if the relaxed-MHD region between the coalescing interfaces is replaced by an ideal-MHD region.  We then present a stable ($k$, pressure) phase space plot, which allows us to determine the form a stable pressure and field profile must take in the region between the interfaces.  From this knowledge, we conclude that there exists a class of single interface plasmas that were found stable by Kaiser and Uecker [R. Kaiser \emph{et al} (2004) \emph{Q. Jl Mech. Appl. Math.}, \textbf{57}, 1], but are shown to be unstable when the interface is resolved.

\end{abstract}

\maketitle

\section{Introduction}\label{sec:intro}

The analysis of a toroidally confined plasma begins with finding solutions to the ideal
	magnetohydrodynamic		%RLD insert
(MHD) equilibrium equation
\begin{equation}\grad P=\BJ\cross\BB,\label{eq:equil}\end{equation}
where $P$ is the scalar pressure, $\BJ=\curl \BB/\mu_0$ the current density, $\mu_0$ the permeability of free space and $\BB$ the magnetic field.  In configurations that have a continuous symmetry
	(which we shall call the 2-D case), such as tokamaks and reversed-field pinches, or two continuous symmetries (the 1-D case), such as large-aspect-ratio tokamaks and reversed-field pinches in the cylindrical approximation, %RLD insert
it is possible to find mathematically rigorous solutions to this equation by assuming the existence of magnetic flux surfaces and using the Grad-Shafranov equation.\citea{ref:Wesson}  However, for the general case of 3-D equilibria, such as stellarators, which have no ignorable spatial coordinate, magnetic flux surfaces are not guaranteed to exist.  Indeed, the most generic situation for field lines in 3-D is that they are chaotic and come arbitrarily close to any point in a volume.\citea{ref:Boozer04}

The problem posed by chaotic fields for the existence of 3-D configurations
was discussed by Grad who noted in 1967 \citea{ref:Grad67}
that flux surfaces are only automatic when an axial symmetry is present.  However, the Kolmogorov--Arnold--Moser
(KAM) theorem \citea{ref:Lichtenberg} shows that flux surfaces with
sufficiently irrational rotational transform, referred to as KAM surfaces,
may survive symmetry breaking perturbations of a given magnetic field.
The situation in a self-consistent MHD equilibrium is more complex, but it is reasonable to conjecture that there exist equilibria consisting of regions in which the magnetic field is partially chaotic separated by perfect flux surfaces called KAM barriers or interfaces.\citea{Hudson_Hole_Dewar_07}

Though regions between the KAM barriers will not in reality be uniformly chaotic, the simplest approach rigorously consistent with both ideal MHD and the existence of embedded chaotic regions is to adopt a ``worst case scenario''  and treat these regions as if they were completely chaotic.  As chaotic fields cannot support pressure gradients, this leads to the \emph{stepped-pressure-profile model} \citea{ref:Hole06,ref:Hole07}, where the pressure and rotational transform steps are positioned at flux surface interfaces and the pressure gradient is zero elsewhere.

In regions of constant pressure, often called force-free regions in astrophysics,
the magnetic field is described by $\curl\BB=\alpha(\Br)\BB$, where $\alpha(\Br)$ is an arbitrary scalar function. By taking the divergence of both sides of this equation and using the vector identities $\dive\curl\BB=0$ and $\dive[\alpha(\Br) \BB]=\alpha(\Br)\dive\BB+\BB\dotv\grad\alpha(\Br)$, we see that
$0=\BB\dotv \grad \alpha(\Br)$ and so the function $\alpha(\Br)$ must be constant along a field line.  Therefore, in chaotic regions, the magnetic field satisfies the equation
\begin{equation}
	\label{eq:Beltrami}
	\curl\BB=\alpha\BB \;,
\end{equation}
with $\alpha$ constant throughout the region, such a field being known as a \emph{Beltrami field} (or linear force-free field). Thus, in the stepped-pressure-profile model the magnetic fields in the constant-pressure regions are assumed to be Beltrami, but with different $\alpha$ in each region.

The Beltrami  equation is the Euler--Lagrange equation for  the relaxed-MHD variational principle of Taylor \citea{Taylor_86} for equilibria of plasmas (such as reversed-field pinches) that undergo a strongly turbulent phase during which all ideal-MHD invariants except for the magnetic helicity and toroidal and poloidal magnetic flux are broken. Thus we shall use the same mathematical framework, the \emph{relaxed-MHD energy principle}, for modelling both plasmas with magnetic-field-line chaos due to 3-D equilibrium effects and turbulent plasma experiments as considered by Taylor. Our results in this paper will in fact be more relevant to the 1-D and 2-D systems considered by Taylor as we shall mainly work in the cylindrical approximation.  The discussion of field-line chaos in 3-D systems above is mainly to motivate our stepped-pressure-profile multi-region relaxation model, but recent work \citea{Tassi_Hastie_Porcelli_07} suggests that multi-region relaxation is relevant to the 2-D reversed-field pinches as well. %end RLD 5/8/08

As a first step towards solving for the field of these plasmas, Hole \emph{et al.} \citea{ref:Hole06,ref:Hole07} considered a periodic cylindrical model.  In this geometry, nested flux surfaces exist everywhere and analytic solutions for the magnetic field exist when the pressure is constant.  Such plasmas form the foundation of this work.

Hole \emph{et al.} used a variational treatment to construct equilibrium solutions and analyse their stability in cylindrical geometry.\citea{ref:Hole07}  Using this analysis, they reproduced the earlier conclusion of Kaiser and Uecker \citea{ref:Kaiser04}, that stable plasmas exist in the case of a single-interface configuration providing there is a jump in the rotational transform, $\ibar$, at the plasma vacuum interface.  However, their work also raised questions.  First, as part of their stability calculations, they compared the stability of a single-interface pressureless plasma with a jump in $\ibar$ to a two-interface plasma
with the same net jump
in the limit that the two interfaces became arbitrarily close.  In doing so they found that the stability was different in the two cases: the two-interface plasma was unstable when the single-interface plasma was not.  Second, the current density must be infinite at any interface where there is a jump in pressure.  Although this is consistent with MHD, it is unphysical in a real plasma, and raises the question of whether the stability conclusions hold if interface is replaced by a small region over which the pressure and field vary continuously.

In this paper, it will be argued that this singular limit paradox is due to the assumption of a relaxed-MHD region between the coalescing interfaces, which allows a tearing instability to occur.
Instead, the single-interface configuration with jump in rotational transform should be compared to a plasma containing a thin
but finite ideal region, with constraints on the helicity of every field line.  If the pressure and magnetic field vary continuously between the values on each side of the interface, the infinite current densities of the stepped pressure model are also removed.

A larger motivation for our study is to remove the singularity in the current density, thereby illuminating some limitations in the stability conclusions of Kaiser and Uecker \citea{ref:Kaiser04}. We will show that if the interface is resolved as an ideal region of nonzero width, the rotational transform profile, $\ibar(r)$, may pass through ideal resonances.  If so, the stability analysis must consider the structure of the pressure profile in the barrier in the limit it is spatially resolved.  In some cases, this analysis will force us to conclude that there is no barrier field and pressure configuration that is both stable and physically plausible, even when Kaiser and Uecker's analysis showed the plasma could support pressure jumps at the interface.

In order to achieve these goals, it is first necessary to develop a stability analysis that can calculate the stability of a plasma consisting of both Taylor-relaxed and ideal-MHD regions.  Section (\ref{sec:stab}), derives a stability test for multiple-interface ideal-MHD plasmas, based on the calculus of variations and section (\ref{sec:cyl}) applies the test to cylindrical plasmas.  Section (\ref{sec:belideal}) describes how the analysis may be modified to incorporate
plasmas with Taylor-relaxed regions.
Section (\ref{sec:twoint}) then describes the illustrative two-interface plasma that will be investigated, and the numerical method used to generate stable solutions and determine their stability.  Finally, section (\ref{sec:vanint}) demonstrates that when the resistance of the narrow plasma region is taken to be zero, there is no discrepancy between the one- and two-interface results. It will also be shown that the stability of a configuration is related to the form of the pressure profile, not just the amount by which it varies between the interfaces.

\section{Ideal- and Relaxed-MHD Multi-Interface Plasmas}\label{sec:stab}
Our MHD model of an $N$-interface plasma draws heavily on two existing approaches.  Recently, Hole \emph{et al.} \citea{ref:Hole06}, developing earlier work by Spies \emph{et al.} \citea{ref:Spies01}
and Kaiser and Uecker \citea{ref:Kaiser04}, determined the stability of an $N$-interface plasma consisting entirely of Taylor-relaxed regions by constructing an $N\times N$ matrix and calculating its eigenvalues.  In 1960, Newcomb \citea{ref:Newcomb60} developed a method to analyse the stability of an ideal cylindrical plasma.  This section will present an extension of the derivation used by Hole \emph{et al.}, using, and comparing with, results of Newcomb where appropriate.

Like Hole \emph{et al.}, our system comprises $N$ nested plasma regions surrounded by a vacuum and enclosed by a perfectly conducting wall, as shown in figure (\ref{fig:mattregions}).  The plasma regions are denoted by $\Pl_i$, with $\Pl_1$ at the core, and $\Pl_N$ next to the vacuum region $\V$.  Each plasma region $\Pl_i$ is bounded on the outer edge by the interface $\I_i$ and on the inner edge by $\I_{i-1}$, while the perfectly conducting wall, $\W$ encases the vacuum.  The interfaces are located at $r=r_i$, the wall at $r=r_w$ and the width of the plasma is normalised so that $r_N=1$.  In the following, we adopt the convention that when equations refer to a particular region or interface, the region or interface will be given before the equation.

\insertfigure{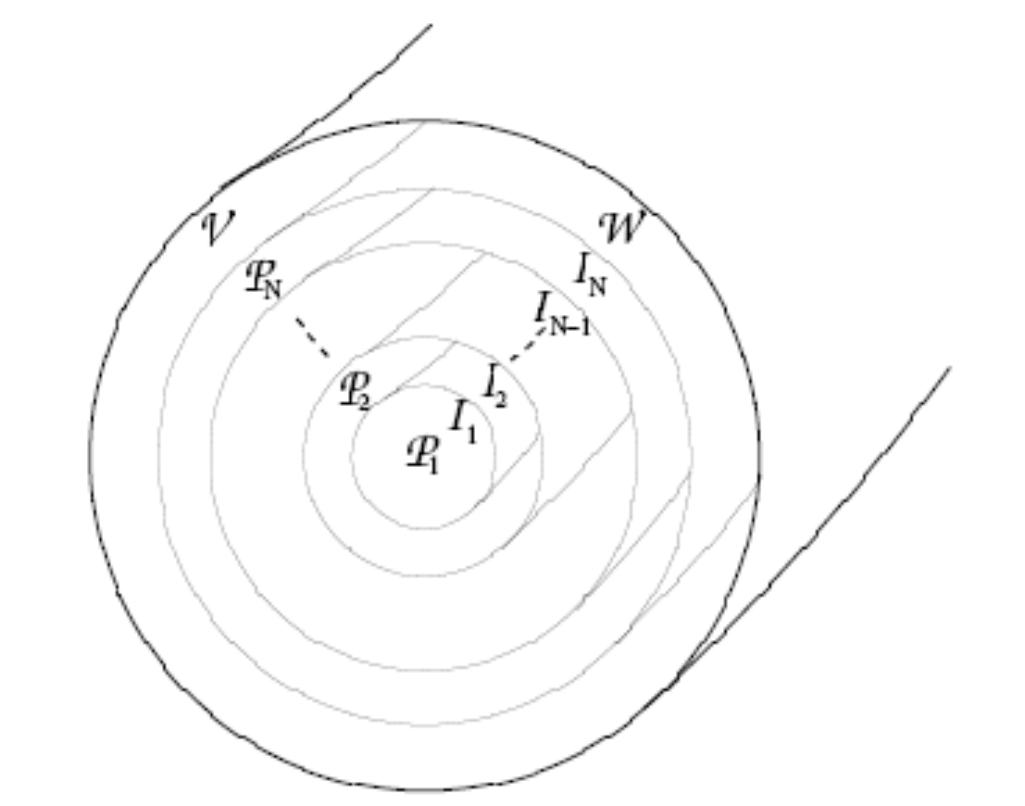}{Diagram of plasma regions.}{mattregions}

We employ a variational approach to study equilibrium and stability.  The potential energy of a $N$-region plasma, where each region is described by \emph{either} ideal or relaxed MHD, can be written
\begin{equation}
	W=\sum_{i=1}^N \int_{\Pl_i} \bigg( \frac{P}{\gamma-1}+\frac{\BB\dotv\BB}{2\mu_0}\bigg)\id\tau^3
	+\int_{\V} \frac{\BB\dotv\BB}{2\mu_0}\id\tau^3,
	\label{eq:potenergy}
\end{equation}
where $P$ and $\BB$ are the pressure and magnetic fields and $\gamma$ is the ratio of specific heats.
We expand $W$ to second order in perturbations from its equilibrium value: $W = W_0 + \delta W +\frac{1}{2}\delta^2 W$, where $\delta W$, the \emph{first variation}, is linear in the perturbation and $\delta^2 W$, the \emph{second variation}, is quadratic (note that we have used the notation of Spies \emph{et al.} \citea{ref:Spies01} where $\delta^2 W$ is twice the quadratic energy functional $\delta W$ used by \citea{Bernstein_etal_58} or $W$ used by Newcomb \citea{ref:Newcomb60}). The system is in equilibrium if the first variation is equal to zero for every possible displacement of the plasma and stable if the second variation is positive for every displacement.

\subsection{Ideal Regions}\label{sec:ideal}
To calculate the variation of this integral in plasma regions described by ideal MHD, we introduce the displacement vector $\Bxi$, defined as the displacement of each fluid element as a function of its original position $\Br_0$.  Eulerian variations in the pressure, $p \equiv \delta P$, and magnetic field, $\Bb \equiv \delta \BB$, can both be related to $\Bxi$, provided it is assumed that the plasma resistivity
and thermal conductivity
are zero, \citea{Bernstein_etal_58,ref:Newcomb62} yielding
\begin{eqnarray}
	p &=&-(\Bxi\cdot\nabla P+\gamma P\nabla\cdot\Bxi)\label{eq:deltaP},
	\\
	\Bb &=&\curl(\Bxi\cross\BB).
	\label{eq:deltaB}
\end{eqnarray}
Equation (\ref{eq:deltaB}) incorporates a continuum of ``frozen-in flux'' constraints \citea{Newcomb_58} holonomically. These constraints do not apply in the vacuum, so in $\cal V$ we use the representation $\Bb = \curl\Ba$, where is a perturbed vector potential, related to $\Bxi$ only at the interfaces through the boundary condition \citea{Bernstein_etal_58}
\begin{equation}
	\label{eq:abc}
	\Bn\cross\Ba = -\Bn\dotv\Bxi\,\BB,
\end{equation}
where $\BB$ is the unperturbed vacuum magnetic field at the interface.

Using these expressions, and simplifying the resulting expression for the variation of a volume integral using vector identities, we find that
\begin{eqnarray}
	\delta W&=&\sum_{i=1}^N\int_{\Pl_i}\Bxi\dotv\left[\grad P
	-\BJ\cross\BB\right] \id\tau^3+\int_{\V}\Ba\dotv\BJ \id\tau^3\nonumber \\
	&&+\sum_{i=1}^N\int_{\I_i}(\Bn\dotv\Bxi)\lbrac P+\frac{B^2}{2\mu_0}\rbrac\id\sigma^2,
	\label{eq:varW}
\end{eqnarray}
where $\Bn$ is the unit outward normal at an interface, $\BJ=\curl\BB/\mu_0$ is the current density, and $\lbrac X\rbrac$ in indicates the change in the quantity $X$ across the interface.
As the condition $\delta W=0$ must be satisfied for all possible $\Bxi$ and $\Ba$, including those that approach a delta function, it is equivalent to
\begin{eqnarray}
	&\Pl_i:& \grad P-\BJ\cross\BB=0,\label{eq:equilplasma}
	\\
	&\V:& \BJ=0,\label{eq:equilvac}
	\\
	&\I_i:&\lbrac P+\frac{B^2}{2\mu_0}\rbrac=0,\label{eq:equilinterface}
\end{eqnarray}

Since the second variation will only be used to determine the stability of equilibrium solutions, it will be assumed that equations (\ref{eq:equilplasma}), (\ref{eq:equilvac}) and (\ref{eq:equilinterface})
are satisfied.  With these assumptions, the second variation is  \citea{Bernstein_etal_58}
\begin{eqnarray}
	\delta^2 W&=&\sum_{i=1}^N\int_{\Pl_i}
	\left[ \frac{\Bb\dotv\Bb}{\mu_0}-\Bxi\cross\BJ\dotv\Bb
	+(\dive\Bxi)(\Bxi\dotv\grad P)+\gamma P (\dive \Bxi)^2 \right]\id\tau^3
	\nonumber\\&&
	+\int_{\V}  \frac{\Bb\dotv\Bb}{\mu_0}\id\tau^3
	+\sum_{i=1}^N\int_{\I_i} (\Bn\dotv\Bxi)^2\lbrac(\Bn\dotv\grad)(P+\frac{B^2}{2\mu_0})\rbrac\id\sigma^2.\label{eq:2ndvar}
\end{eqnarray}
If $\delta^2 W$ is positive for all physical perturbations, the configuration is stable.  Finding the global minimum of $\delta^2 W$ is a non-trivial problem. Instead, as is done elsewhere \citea{ref:Newcomb60,ref:Hole06} a vector equation will be found which is satisfied by any perturbation for which $\delta^2 W$ is a local minimum or maximum.

\subsection{Relaxed and Vacuum Regions}\label{sec:vacrelax}

As relaxed-MHD regions are by their nature strongly mixed, fluid displacements are ill-defined in these regions. Nevertheless, in the following we discuss how to formulate an energy principle in these regions that is as close as possible to that used in ideal-MHD regions. As we can regard the vacuum field as a Beltrami field for which $\alpha  = 0$, the discussion applies to the vacuum region as well, where \emph{a fortiori} $\Bxi$ cannot be thought of as a fluid displacement.

The fundamental representation of the perturbed magnetic field in these regions is $\Bb = \curl\Ba$ with boundary condition \eqn{abc}. As \eqn{deltaB} makes it clear that we can take $\Ba = \Bxi\cross\BB$ throughout an ideal-MHD region, a unified representation for the perturbed magnetic field is provided by using a gauge such that
\begin{equation}
	\label{eq:aNewc}
	\Ba=\Bxi\cross\BB,
\end{equation}
which is consistent with the boundary condition \eqn{abc} at the edge of a relaxed or vacuum region and effectively \emph{defines} the projection of $\Bxi$ perpendicular to $\BB$ throughout the plasma and vacuum regions (the parallel component being arbitrary). This gauge was introduced by Newcomb \citea{ref:Newcomb60} for vacuum fields and we shall henceforth refer to it as the \emph{Newcomb gauge}.

Newcomb noted that $\Bxi$ can be singular in  this gauge. To see this, let $s$ be a coordinate (e.g. $r$ in cylindrical geometry) labelling magnetic surfaces. Then $\Bb\dotv\grad s \equiv \BB\dotv\grad (\Bxi\dotv\grad s)$. This can be inverted to give $\Bxi\dotv\grad s = (\BB\dotv\grad)^{-1}(\Bb\dotv\grad s)$ except at magnetic surfaces where
$\BB\dotv\grad$ is singular (within the function space of trial functions we allow). If there is such a singular surface within the region under consideration, then $\Bxi$ will diverge in the neighbourhood of the singular surface. This will be made more explicit when we consider the cylindrical case in \Sec{belideal}. %RLD 14/8/08

Provided we treat these singularities appropriately, the Newcomb gauge will allow us to unify the discussion of ideal, relaxed and vacuum regions. However, before discussing the application of the Newcomb gauge to relaxed MHD we first review the standard approach of using an arbitrary gauge.

The multiregion \citea{Hudson_Hole_Dewar_07,Dewar_etal_08} generalization of the relaxed-MHD energy principle is to minimize the total potential energy  $W$ \eqn{potenergy} under the constraints of constant magnetic helicity
\begin{equation}
	\label{eq:Helicity}
	K_i \equiv \frac{1}{2\mu_0}\int_{{\cal P}_i} \BA\dotv \BB \id\tau^3,
\end{equation}
and constant $P_iV_i^{\gamma}$ in each relaxed region, of volume $V_i$, where $P = P_i$ is constant throughout the region (i.e. $\grad P \equiv 0$ in ${\cal P}_i$). Note that we have defined $K$ with a prefactor of $1/2\mu_0$, so the helicity as commonly defined \citea{Taylor_86} is $2\mu_0 K$.

The pressure constraints can be implemented holonomically \citea{Dewar_etal_08}, while the helicity constraints are implemented by  introducing Lagrange multipliers $\alpha_i$, and replacing the constrained problem $\delta W = 0$ with that of finding an extremum of a magnetic ``free energy'' \citea{Bhattacharjee_Dewar_82},
\begin{equation}
		\label{eq:FreeEn}
	F =  W - \sum_{i=1}^N \alpha_i K_i \;,
\end{equation}
subject only to the constraint that variations $\Ba$ not vanishing on the boundary satisfy \eqn{abc}.

The first variation is \citea{ref:Spies01,Dewar_etal_08}
\begin{eqnarray}
	\delta F&=&\sum_{i=1}^{N+1}\int_{\Pl_i}\Ba\dotv(\curl\BB - \alpha_i\BB) \id\tau^3
	\\
	&&+\sum_{i=1}^N\int_{\I_i}(\Bn\dotv\Bxi)\lbrac P+\frac{B^2}{2\mu_0}\rbrac\id\sigma^2.
	\label{eq:varF}
\end{eqnarray}
Setting $\delta F=0$ for all possible $\Ba$ and $\Bxi$, we have the equilibrium equations
\begin{eqnarray}
	&\Pl_i:& \curl\BB=\alpha_i\BB, \label{eq:equilplasmaR}
	\\
	&\I_i:&\lbrac P+\frac{B^2}{2\mu_0}\rbrac=0,\label{eq:equilinterfaceR}
\end{eqnarray}
\eqn{equilplasmaR} being a Beltrami equation \eqn{Beltrami}  in each relaxed region, with $\alpha_{N+1} \equiv 0$.

It can also be shown for Beltrami fields (cf. \citea{ref:Spies01} and provided the trial function space of $\Bxi$ is such that $\delta\alpha_i = 0$ and $\delta V_i = 0$ \citea{Dewar_etal_08}) that the constrained second variation $\delta^2 W$ is the same as the unconstrained second variation $\delta^2 F$
\begin{eqnarray}
	\delta^2 F&=&\sum_{i=1}^{N+1}\int_{\Pl_i}
	\frac{1}{\mu_0}(\Bb\dotv\Bb - \alpha_i\Ba\dotv\Bb) \id\tau^3
	\nonumber\\&&
	+\sum_{i=1}^N\int_{\I_i} (\Bn\dotv\Bxi)^2\lbrac(\Bn\dotv\grad)\frac{B^2}{2\mu_0}\rbrac\id\sigma^2.\label{eq:2ndvarR}
\end{eqnarray}

We now compare the relaxed-MHD and ideal-MHD variations, $\delta^n F$ and $\delta^n W$ respectively,  to show they are equivalent in the Newcomb gauge \eqn{aNewc} when evaluated on a relaxed equilibrium, i.e. when the Beltrami equations \eqn{equilplasmaR} are satisfied. With this assumption the ideal first variation \eqn{varW} implies $\grad P = 0$ and the interface force balance equations \eqn{equilinterfaceR}, as expected. The assumption $\delta V_i = 0$ allows us to set $\dive\Bxi = 0$ (see \Sec{cyl}). Then, recognizing that $\Bxi\cross\BJ\dotv\Bb = \alpha_i \Ba\dotv\Bb/\mu_0$, we see that \eqn{2ndvar} is indeed the same as \eqn{2ndvarR} for a Beltrami equilibrium. Because of this equivalence we shall simply take it as understood that $\delta^2 W$ means $\delta^2 F$ in a relaxed-MHD region.

\subsection{Kinetic Energy Normalisations}\label{sec:kinnorm}

When searching for local minima of $\delta^2W$, two problems are immediately encountered.

First, since $\delta^2 W$ scales with the square of the amplitude of $\Bxi$, a finite minimum need not exist (and, if it does, it is the trivial minimum $\delta^2W = 0$ attained when $\Bxi = 0$). %RLD 18/8/08
To compensate, we normalise $\delta^2 W$
by assigning a nonzero value to a positive-definite quadratic functional, $\norm[\Bxi]$.
For the purpose of determining stability from the energy principle, the choice of normalisation is not unique; apart from being quadratic and positive it is necessary only that $\norm$ be finite for all admissible functions $\Bxi$.
In this section we discuss criteria for choosing $\norm$, interpret it in terms of a kinetic energy, and make a simple choice for use in the following sections.

The second problem is that the values of $\lambda=\delta^2 W/\norm$ may approach a lower bound that is not attained by any physical perturbation.  The class of perturbations considered will therefore be extended to a \emph{generalised function} space that
includes any vector function with the following properties:
\begin{enumerate}
\item The function may be approximated to any degree of accuracy by a physical perturbation.
\item The value of $\delta^2 W$ for the vector function is finite and approached by the values of $\delta^2 W$ for approximating perturbations.
\end{enumerate}
If $\delta ^2 W$ is positive for all $\Bxi$ satisfying these requirements, then it must be positive for all physical perturbations, and so the configuration is stable.  If a non-physical perturbation exists for which $\delta^2 W$ is negative, the second requirement dictates that there also exists a physical perturbation with negative $\delta^2 W$.\citea{ref:Newcomb60}

To find the minimum value of $\delta^2 W$, subject to the constraint of fixed $\norm$, the technique of Lagrange multipliers is used, setting $\delta L=0$, where the \emph{Lagrangian} $L \equiv \delta^2 W-\nu \norm$.\citea{ref:Spies01}  As $L$ is a symmetric quadratic form with respect to $\Bxi$ \citea{ref:Kulsrud64}, if $\delta L=0$, then $L=0$.  Therefore, when $\delta L=0$, the value of $\nu$ is the same as the value of $\lambda=\delta^2 W/\norm$.  The stability of the configuration can therefore be determined from the sign of the Lagrange multiplier, which will hereafter be referred to as $\lambda$.

A positive eigenvalue can be interpreted as the square of the frequency of a normal mode,  $\lambda = \omega^2$, while a negative eigenvalue, which implies instability, is the square of the growth rate, $\lambda = -\gamma^2$. In this interpretation $\frac{1}{2}\lambda\norm$ is a model kinetic energy. As it chosen for mathematical convenience, $\omega$/$\gamma$ is not necessarily an accurate estimate of a physical frequency/growthrate.

Several choices for the model kinetic energy are reviewed in \citea{Dewar_97b}. In ideal MHD we can in fact normalise with the physical kinetic energy factor $\norm = \int\!\rho\,|\Bxi|^2 \id\tau^3$ where $\rho$ is the mass density. However, to remain closer to Newcomb's \citea{ref:Newcomb60} stability analysis it is convenient to replace $\rho$ with an anisotropic tensor density $\Brho = \rho\,\Bn\Bn$, so that, in an ideal-MHD region ${\cal P}_i$, we use
\begin{equation}
	\label{eq:Newcnorm}
	\norm_{{\rm N},i}[\Bxi] \equiv \int_{{\cal P}_i} \rho \, (\Bn\dotv\Bxi)^2 \id\tau^3,
\end{equation}
which has been used for instance in the PEST2 code \citea{Manickam_Grimm_Dewar_81,Grimm_Dewar_Manickam_83}. This defines not only a normalisation but a \emph{norm} for the function space in which admissible solutions in Newcomb's formulation exist. In this space the linearised force operator is self adjoint and the integration by parts required to obtain Newcomb's ideal-MHD 1-D scalar energy principle, Eq.~(15) of \citea{ref:Newcomb60}, or Bineau's 2-D generalisation of it \citea{Bineau_1962,Manickam_Grimm_Dewar_81,Grimm_Dewar_Manickam_83,Dewar_Pletzer_90}, can be carried out.

In the case of relaxed MHD and the vacuum, $\partial_t\Bxi$ cannot be interpreted as a physical velocity and $\norm_{\rm N}$ is not a physically natural norm. In fact, as it is crucial that the helicity be finite in relaxed MHD the natural norm is one based on the helicity. Thus we define the \emph{helicity norm} for use in a relaxed-MHD or vacuum region
\begin{equation}
	\label{eq:Helnorm}
	\norm_{{\rm H},i}[\Bxi]  \equiv |\int_{{\cal P}_i} h \, \Bxi\cross\BB\dotv\curl(\Bxi\cross\BB) \id\tau^3|
\end{equation}
where $h$ is an arbitrary factor.

Choosing $h$ so $\norm_{{\rm H},i}$ has the same dimensions as $\norm_{{\rm N},i}$ we can define the total normalising factor $\norm$ as a sum over the regions, using $\norm_{{\rm N},i}$ or $\norm_{{\rm H},i}$ as appropriate. However, \emph{provided} the plasma is stable against perturbations that do not move the interfaces, we may effect  further simplification by taking $\rho$ and $h$ to be Dirac delta functions concentrated on the interfaces and use
\begin{equation}
	 \norm=\sum_{i=1}^N\int_{\I_i}d\sigma^2(\Bxi\dotv\Bn)^2.
	\label{eq:norm}
\end{equation}
(If the above assumption is violated then this normalisation will give $\lambda = -\infty$.)

In the following we use \eqn{norm} for normalisation and, with $\rho = 1$ and $h = 1$, use $\norm_{{\rm N},i}$ and $\norm_{{\rm H},i}$ as norms to select the appropriate function space for $\Bxi$ in each region: The ``Newcomb'' norm, $\norm_{{\rm N},i}$, restricts the trial function space for $\Bn\dotv\Bxi$ to square-integrable generalised functions, while the helicity norm, $\norm_{{\rm H},i}$, allows more divergent functions. This will be made more explicit in \Sec{cyl}.% RLD 19/8/08

In evaluating $\delta L$, we consider the equilibrium quantities as fixed, varying $\Bxi$ only.  Using vector identities and equation (\ref{eq:2ndvar}), we find
\begin{eqnarray}
	\delta L&=&\sum_{i=1}^N\int_{\Pl_i}  \delta\Bxi\dotv\left\{\BB\cross\Bj-\BJ\cross\Bb
	-\grad(\Bxi\dotv\grad P)-\grad[\gamma P (\dive \Bxi)]\right\}\id\tau^3
	+\int_{\V} \delta\Bxi\dotv\BB\cross\Bj\id\tau^3\nonumber
	\\
	&&+\sum_{i=1}^N\int_{\I_i}(\Bn\dotv\delta\Bxi)\left( \frac{\lbrac\BB\dotv\Bb
	 +(\Bxi\dotv\grad)(B^2)\rbrac}{\mu_0}-\lambda(\Bxi\dotv\Bn)
	\right)\id\sigma^2,
	\label{eq:deltaLwithident}
\end{eqnarray}
where $\Bj=\curl\Bb/\mu_0$.  Since $\delta L=0$ for any possible perturbation, including those that approach delta functions, we may conclude that
\begin{eqnarray}&\Pl_i:&\BB\cross\Bj-\BJ\cross\Bb-\grad(\Bxi\dotv\grad P)-\grad[\gamma P (\dive \xi)]=0,\label{eq:vol}
\\&\I_i:&\lbrac(\Bxi\dotv\grad)\frac{B^2}{2\mu_0}\rbrac
+\lbrac\frac{\BB\dotv\Bb}{\mu_0}\rbrac
-\lambda(\Bxi\dotv\Bn)=0\label{eq:surface}
\\&\V:&\curl\Bb=0.\label{eq:volvac}\end{eqnarray}
Equation (\ref{eq:vol}) is a system of three second order partial differential equations, and therefore does not have a unique solution.  In order to proceed further with this analysis, it is necessary to specify a geometry for the configuration.

\section{Cylindrical
ideal-MHD Multi-Interface %RLD 22/8/08
Plasmas}\label{sec:cyl}

\subsection{Solutions in Case of no Singular Surfaces}\label{eq:idealcontinuation}

This section will apply the stability analysis to periodic 1D cylindrical plasmas with coordinates $(r,\theta,z)$, in which the magnetic field and pressure are functions of only the radial coordinate, $r$.  By imposing a periodicity constraint of length $L$ in the $z$ direction, the model is a simple analogue of a large-aspect-ratio, toroidally symmetric plasma with circular flux surfaces.  In this analogue, $\theta$ and $2\pi z/L$ correspond to the poloidal and toroidal angles, and $L=2 \pi R$, with $R$ the major radius of the toroidal plasma.

Like Hole \emph{et al.} \citea{ref:Hole06}, we next decompose $\Bxi$ into a Fourier series
with components:
\begin{equation}\Bxi_{m,k}(r,\theta,z)=[\xi_{m,k,r}(r)\rhat+\xi_{m,k,\theta}(r)\thhat+\xi_{m,k,z}(r)\zhat]\exp[i(m\theta+k z)].\label{eq:Fouriercomp}\end{equation}
Here, the first two subscripts refer to the wave number in the $\theta$ and $z$ directions respectively, the third subscript refers to the geometric component, $\rhat$, $\thhat$ and $\zhat$ are unit vectors, $k=-2\pi n/L$, and $m$ and $n$ are integers.  The second variation is decomposed as
a sum of uncoupled terms
\begin{equation}\delta^2 W=\sum_{m,k}\delta^2 W_{m,k},\label{eq:deltaWmk}\end{equation}
where $\delta^2 W_{m,k}$ is the second variation evaluated for the $m,k$ component.
Thus we can consider each Fourier component individually, restricting the trial function space to a single value of $m$ and $k$, so that \eqn{vol} reduces to a second order linear ODE.

Under the Fourier decomposition of equation (\ref{eq:Fouriercomp}), the dot product of equation (\ref{eq:vol}) with $\BB$ reduces to
\begin{equation}\BB\dotv\grad [\gamma P(\dive\Bxi_{m,k})]=0,\label{eq:divonlines}\end{equation}
where we have used $\BB\dotv\grad P=0$ and equation (\ref{eq:equilplasma}).  The quantity $\gamma P(\dive\Bxi_{m,k})$ is therefore constant along field lines, and by extension on magnetic surfaces in the cylinder.  As the pressure is constant along magnetic surfaces, $\dive\Bxi_{m,k}$ must also be constant.  However,
$\dive\Bxi_{m,k}$ is a function of $r$ multiplied by $\exp[i(m\theta+k z)]$
and so if the divergence is to be constant on a magnetic surface, $\dive\Bxi_{m,k}=0$, except if $m=k=0$.
Therefore, ignoring the case $m=k=0$, the normal modes $\Bxi_{m,k}$ must satisfy
\begin{eqnarray}&\Pl_i:&\BB\times\Bj_{m,k}-\BJ\times\Bb_{m,k}-\grad[\Bxi_{m,k}\dotv\grad P]=0,\label{eq:volagain}
\\&&\dive\Bxi_{m,k}=0,\label{eq:divxiagain}
\\&\I_i:&\lbrac(\Bxi_{m,k}\dotv\grad)\frac{B^2}{2\mu_0}\rbrac
+\lbrac\frac{\BB\dotv\Bb_{m,k}}{\mu_0}\rbrac
-\lambda(\Bxi_{m,k}(r)\dotv\Bn)=0,\label{eq:surfaceagain}\end{eqnarray}
where $\Bb_{m,k}=\curl(\Bxi_{m,k}\cross\BB)$ and $\Bj_{m,k}=\curl \Bb_{m,k}$.

First we solve for $\Bxi_{m,k}$ in the plasma.  Using the $\theta$ component of equation (\ref{eq:volagain}), along with the requirement $\Bxi_{m,k}$ is divergence free, we are able to eliminate $\xi_{m,k,\theta}(r)$ and $\xi_{m,k,z}(r)$, yielding
\begin{equation}\frac{d}{dr}\left(f(r) \frac{d\xi_{m,k,r}(r)}{dr}\right)-g(r)\xi_{m,k,r}(r)=0,\label{eq:NewcombEL}\end{equation}
where
\begin{eqnarray}f(r)&=&\frac{r[m\Bth(r)+k r \Bz(r)]^2}{k^2r^2+m^2},\label{eq:Newcombf}
\\g(r)&=&\frac{[m\Bth(r)-k r \Bz(r)]^2}{r(k^2r^2+m^2)}+\frac{[m\Bth(r)+k r \Bz(r)]^2}{r}-\frac{2\Bth(r)}{r}\frac{d}{dr}[r\Bth(r)]\nonumber
\\&&+\frac{d}{dr}\left(\frac{m^2\Bth(r)^2-k^2 r^2\Bz(r)^2}{k^2r^2+m^2}\right).\label{eq:Newcombg}\end{eqnarray}
This is equation (23) of Newcomb \citea{ref:Newcomb60}, which was derived by first eliminating $\xi_{m,k,\theta}(r)$, $\xi_{m,k,z}(r)$ and then taking the variation, rather than the opposite order presented here.  Newcomb uses equation (\ref{eq:equilplasma}) to rewrite $g(r)$ as
\begin{eqnarray}g(r)&=&\frac{2k^2r^2}{k^2r^2+m^2}\frac{dP}{dr}+\frac{[m\Bth(r)+k r \Bz(r)]^2(k^2r^2+m^2-1)}{r(k^2r^2+m^2)}\nonumber
\\&&+\frac{2k^2r[k^2r^2\Bz(r)^2-m^2\Bth(r)^2]}{(k^2r^2+m^2)^2}.\label{eq:Newcombg2}\end{eqnarray}
Equation (\ref{eq:NewcombEL}) has a singular point at $r=0$ and whenever
\begin{equation}k r \Bz(r) + m \Bth(r)=0,\label{eq:rsing}\end{equation}
or equivalently
\begin{equation}\frac{k}{m}+\frac{\ibar(r)}{R}=0,\label{eq:rsingibar}\end{equation}
where $\ibar(r)=\iota(r)/(2\pi)$ is the rotational transform, defined in cylindrical geometry as
\begin{equation}\ibar(r)=\frac{R\Bth(r)}{r\Bz(r)}.\label{eq:secibardef}\end{equation}
Surfaces corresponding to values of $r$ at which equation (\ref{eq:rsing}) is satisfied are also known as resonant surfaces since on these surfaces the phase of the displacement is constant along magnetic field lines, i.e. $\Bk\dotv\BB=0$, where $\Bk$ is the wave vector.
It is at these surfaces that the operator $\BB\dotv\grad$ discussed in \Sec{vacrelax} is singular in the trial function space $m,k$. %RLD21/8/08
Initially it will be assumed that there are no solutions to equation (\ref{eq:rsing}) in a given region.  The case when singular points are present will be treated separately.

Next, we choose a representation for $\Bxi_{m,k}$ on an interface ${\cal I}_i$.  To evaluate the interface terms, $\Bb_{m,k}$ is expanded in equation (\ref{eq:surfaceagain}) and the expressions for $\xiz(r)$ and $\xith(r)$ found when solving equation (\ref{eq:vol}) are substituted in, yielding
\begin{equation}
	-\frac{\lbrac k^2 r \left[\Bz(r)^2+\Bth(r)^2\right] \xir(r)\rbrac_i}{m^2+k^2 r_i^2}
	-\frac{ \lbrac f(r) \xi\der{r}(r)\rbrac_i}{r_i}-\lambda\xir(r_i)=0
	\label{eq:surfacexir}
\end{equation}
Since equation (\ref{eq:NewcombEL}) has two linearly independent solutions, it is necessary to find boundary conditions that allow easy evaluation of equation (\ref{eq:surfacexir}), ultimately by converting the $N$ equations to a matrix.
Let $X_i$ be the value of $\xi_{m,k,r}(r)$ on the $i$th interface, which is located at $r=r_i$.  Ignoring for the moment the first plasma region and the vacuum, the solution to equation (\ref{eq:NewcombEL}) that satisfies the boundary conditions
\begin{eqnarray}
	 \xi_{m,k,r}(r_{i-1})=X_{i-1},&&\xi_{m,k,r}(r_{i})=X_{i},\label{eq:boundaryEL}
\end{eqnarray}
may be written as
\begin{equation}
	 \xi_{m,k,r}(r)=X_{i-1}\zeta_{i,1}(r)+X_{i}\zeta_{i,2}(r),\quad r_{i-1} \leq r \leq r_i, \label{eq:basisEL}
\end{equation}
where $\zeta_{i,1}{r}$ and $\zeta_{i,2}(r)$ are
two independent basis
solutions that satisfy the boundary conditions
\begin{eqnarray}\zeta_{i,1}(r_{i-1})=1&&\zeta_{i,1}(r_{i})=0,\nonumber
\\ \zeta_{i,2}(r_{i-1})=0&&\zeta_{i,2}(r_{i})=1.\label{eq:basisboundaryEL}\end{eqnarray}
Substituting
\eqn{basisEL}
and its derivative into equation (\ref{eq:surfacexir}), we see that the interface equations depend linearly on the $X_i$.  In the case of the outermost interface, the requirement on the wall $\W:\Bn\dotv\Bb=0$ provides the boundary condition
% $\zeta_V(r_w)=0$
$\xi_{m,k,r}(r_w)=0$
and the
surviving basis solution $\zeta_{N+1}$
is normalised to 1 at $r_N$.  In the first plasma region, the
single basis
solution is chosen so that it remains finite at the singular point $r=0$ and is normalised to 1 at $r_1$.  This solution satisfies the boundary conditions \citea{ref:Newcomb60}
 \begin{equation}\left.\begin{array}{cc} \zeta_{1}(0)=0, & |m|\ne1
 \\ \zeta\der{1}(0)=0, &|m|=1\end{array}\right..\label{eq:r0boundcondxir}\end{equation}
Using the solutions $\zeta_{i,1}$ and $\zeta_{i,2}$, the
$N$ interface equations given by equation (\ref{eq:surfacexir}) may be written in matrix form
\begin{equation}
	A\BX=\lambda\BX\label{eq:matrix}
\end{equation}
where
\begin{equation}
	\BX=\{X_1,\ldots,X_N\}.\label{eq:interfacevector}
\end{equation}
The stability of the $m,k$ component of $\Bxi$ can then be confirmed by checking that all eigenvalues of % $M$
the $N\times N$ matrix $A$ are positive.

\subsection{Ideal Solutions with Singular Surfaces}\label{eq:idealcontinuation}

Up to this point, it has been assumed there are no resonant surfaces in the plasma region $\Pl_i$.
Under this assumption
the solutions to equation (\ref{eq:NewcombEL}) will be finite and continuous,
corresponding
to physical perturbations.
However, if
there is a resonant surface in the region, the solutions are not guaranteed to be finite, and so it is necessary to check that,
in an ideal-MHD region,
the solutions meet the two conditions for test functions given in section (\ref{sec:stab}): the function may be approximated to any degree of accuracy by a physical perturbation and the value of $\delta^2 W$ for the vector function is finite and approached by the values of $\delta^2 W$ for approximating perturbations.

Let us assume there exists one singular point $r_s$ in $\Pl_i$, that satisfies equation (\ref{eq:rsing}).  By taking a series expansion of the functions $f(r)$ and $g(r)$ in equation (\ref{eq:NewcombEL}) about $r_s$ subject to the condition that equation (\ref{eq:rsing}) is satisfied, it can be shown that,
to leading order in $r - r_s$,
$f(r)=\alpha (r-r_s)^2$ and $g(r)=\beta$ \citea{ref:Newcomb60}, where
\begin{eqnarray}\alpha&=&\frac{r_s}{k^2r_s^2+m^2}\bigg(kr_sB\der{z}(r_s)+k\Bz(r_s)+mB\der{\theta}(r_s)\bigg)^2,\label{eq:fapprox}
\\\beta&=&\frac{2k^2r_s^2}{k^2r_s^2+m^2}P^\prime(r_s).\label{eq:approxg}\end{eqnarray}
Making the change of variable $x=r-r_s$, equation (\ref{eq:NewcombEL}) may then be approximated by
\begin{equation}\alpha\frac{d}{dx}\bigg(x^2\frac{d\xir(x)}{dx}\bigg)-\beta\xir(x)=0.\label{eq:approxEL}\end{equation}
The solutions to this equation have the functional form $|x|^\nsing$, where
\begin{equation}\nsing=\frac{-1\pm\sqrt{1+4\beta/\alpha}}{2}.\label{eq:approxsoln}\end{equation}
These solutions form the leading terms of the full Frobenius expansion of $\xi_r$ about the singular point. If the two $\eta$ roots differ by an integer, the full expansion may include logarithmic terms.\citea{ref:Ince}  This is particularly relevant to this work, because when $P\der{}(r)=0$ the two solutions to equation (\ref{eq:approxsoln}) are $0$ and $-1$.  It turns out that in the special case of a Beltrami field the logarithmic terms vanish, however this is not the case for constant pressure plasmas in general.

If $\alpha+4\beta<0$, the two values of $\nsing$ are complex, and so the solution oscillates.  It is also the case that the frequency of oscillation increases as the singular point is approached.  Such solutions are associated with instability \citea{ref:Newcomb60}, \citea{ref:Freidberg} and so a first requirement for stability is that $\alpha+4\beta>0$.  Using the definition of $\ibar$ and the resonance condition, the inequality may be reduced to Suydam's condition
\begin{equation}\frac{r \Bz(r)^2}{8 R^2}\left(\frac{d\log(\ibar)}{dr}\right)^2+\frac{dP}{dr}>0.\label{eq:Suydam}\end{equation}
If Suydam's condition is satisfied, the two values of $\nsing$ are real.
One of the solutions, known as the \emph{small solution},  $\nsing_s > -1/2$,  diverges slower than $1/(r - r_s)^{1/2}$ as $r \rightarrow r_s$, the other, known as the \emph{large solution}, $\nsing_l<-1/2$, diverging faster than $1/(r - r_s)^{1/2}$.
Because the singular point at $r_s$ is excluded, these solutions are not ordinary point wise solutions of Newcomb's Euler--Lagrange equation, but are better regarded as generalised functions \citea{Dewar_Persson_93}. From this point of view there are actually \emph{four} solutions in the neighbourhood of $r_s$: two small solutions and two large solutions, depending on whether they have support on one side or the other of $r_s$ (or, equivalently, small and large solutions that are odd and even about $r_s$ to leading order). However, the two large solutions are excluded in ideal MHD because the norm $\norm_{{\rm N},i}$ defined in \eqn{Newcnorm} is infinite. Thus, in ideal MHD we still have only two independent solutions as in the non-singular case, but they are discontinuous at $r_s$, as noted by Newcomb \citea{ref:Newcomb60}.

Using a similar construction to that used when no singular points are present, two
basis
solutions to equation (\ref{eq:NewcombEL}) are defined according to the boundary conditions
\begin{eqnarray}
	\zeta_{i,1,S}(r_{i-1})=1,
	&\lim_{r\to r_s^-}\zeta_{i,1,S}(r)=a|r-r_s|^{\nsing_s},
	& \zeta_{i,1,S}(r) = 0 \:\: \mbox{for} \:\: r\in(r_s,r_i],
	\\
	\zeta_{i,2,S}(r_{i})=1,
	&\lim_{r\to r_s^+}\zeta_{i,2,S}(r)=b|r-r_s|^
	 {\nsing_s},&\zeta_{i,2,S}(r) = 0 \:\: \mbox{for} \:\:  r\in[r_{s-1},r_s),\nonumber\label{eq:basisboundaryELsing}
\end{eqnarray}
where $\nsing_s$ is the value of $\nsing$ corresponding to the small solution and $a$ and $b$ are constants determined by the interface boundary condition.  As before, the subscript $i$ denotes the plasma region, and the second subscript distinguishes the two independent solutions.  The third subscript indicates the presence of resonant surfaces.
Equation~(\ref{eq:basisEL}) still applies, but with $\zeta_{i,1}$ and $\zeta_{i,2}$ replaced by $\zeta_{i,1,S}$ and $\zeta_{i,2,S}$, respectively.
In contrast to the case where there are no resonant surfaces, where $\zeta\der{i}(r_i)$ is a function of $X_{i-1}$, $X_i$ and $X_{i+1}$, $\zeta\der{i,S}(r_i)$ depends only on $X_i$.  As a consequence, two adjacent off-diagonal entries in the tri-diagonal matrix
$A$ are removed, allowing it to be separated into two smaller matrices for which the eigenvalues may be calculated separately.

Finally, if a singular point exists in the inner plasma region, the boundary condition that $\zeta_{1}(r)$ is finite at $r=0$ is replaced by the requirement that the solution is small
as $r \rightarrow r_s$ from above and vanishes in the interval $0 \leq r < r_s$.
This modification alters the eigenvalues of the matrix
$A$, but not its structure.

\section{Taylor-Relaxed and Ideal-MHD Multi-Interface Plasmas in Cylindrical Geometry}\label{sec:belideal}

This section will describe how to adapt the analysis to plasmas consisting of both ideal regions, in which the analysis of the previous
section is applied, and Taylor-relaxed regions, where the analysis of Hole \emph{et al.} \citea{ref:Hole06} is used.  The analysis of Hole \emph{et al.}
may only be applied to Beltrami fields, where $\grad P=0$ and $\curl\BB=\alpha_i\BB$.  The
stability analysis developed in this paper makes no assumption on the form of the equilibria
in ideal-MHD regions, which can be either Beltrami states or regions with finite pressure gradients.

Suppose the
equilibrium
field in $\Pl_i$ is Beltrami. Equation (\ref{eq:vol}) reduces to
\begin{equation}\Pl_i:\frac{\BB\cross[\curl\Bb_{m,k}]}{\mu_0}-\frac{\alpha_i\BB\times\Bb_{m,k}}{\mu_0}=0,\label{eq:volbel}\end{equation}
where $\alpha_i$ is the Lagrange multiplier of the Beltrami field.  Since this equation does not involve the perturbation $\Bxi$, it is possible to solve for $\Bb_{m,k}$ directly
	\begin{equation}
	\Pl_i:\curl\Bb_{m,k}=\alpha_i\Bb_{m,k}
	+ \alpha_{m,k}\BB , % RLD 23/8/08
	\label{eq:volbel2}
\end{equation}
where $\alpha_{m,k}$ has the same $\theta$ and $z$ dependence as $\Bb_{m,k}$, i.e. $\exp[i(m\theta+k z)]$. Its $r$ dependence is determined by taking the divergence of both sides of \eqn{volbel2}, giving $\BB\dotv\grad\alpha_{m,k} \equiv i(mB_{\theta}/r + kB_z)\alpha_{m,k} = 0$. Thus $\alpha_{m,k}$ vanishes except in the neighbourhood of a singular surface, where it can have the form of a Dirac delta function with arbitrary amplitude, corresponding to a \emph{current sheet} at the singular surface.

Apart from the current sheet term, \eqn{volbel2}
is the same as the
relaxed-MHD
volume equation used by Hole \emph{et al.}  Put another way, $\alpha_{m,k} = 0$ in relaxed MHD. Thus, using the normalisation \eqn{norm}, the existence of current sheets at resonant surfaces is the only feature that distinguishes linearised relaxed-MHD perturbations from ideal-MHD perturbations on a Beltrami equilibrium region (as expected from our general discussion in \Sec{stab}). It is the current sheet that allows the small solutions on either side of a singular surface to be disconnected in ideal MHD, screening one side from the other. In ideal MHD, field-line reconnection is forbidden by the frozen-in flux condition, so the current sheet must form to prevent the tearing-mode island (of arbitrarily small amplitude in the linearised approximation) that forms in relaxed MHD \citea{Taylor_86}.

In order to modify the ideal analysis of \Sec{cyl} to include Beltrami regions, it is only necessary to change the boundary conditions used to define the solutions $\zeta\der{i,1,S}(r)$ and $\zeta\der{i,2,S}(r)$.  Although the perturbation function
$\xi_{m,k,r}$ has no physical meaning in the Taylor analysis, we extend the definition for mathematical convenience
as in \eqn{aNewc}.
Using the
identity $\Bb\dotv\grad r = \BB\dotv\grad (\Bxi\dotv\grad r)$, the radial components $b_{m,k,r}(r)$ and $\xi_{m,k,r}(r)$ are related by
\begin{equation}
	\xi_{m,k,r}(r)=\frac{-ib_{m,k,r}(r)}{m\Bth(r)/r+k \Bz(r)}.\label{eq:xirbrrel}
\end{equation}
As the $r$-component of $\Bb$ must be continuous in $r$ for it to be possible to satisfy $\dive\Bb = 0$, $b_{m,k,r}(r)$ is an even function in the neighbourhood of a singular surface (i.e. asymptotically as $r \rightarrow r_s$). Thus \eqn{xirbrrel} implies that, for relaxed-MHD perturbations, only the asymptotically-odd large $\xi_{m,k,r}$ solution and the asymptotically-even small solution are allowed. (This is consistent with the helicity norm, \eqn{Helnorm}, defined as a principal part integral, as the odd large solution has finite norm but the even large solution does not.)

Thus, as in the ideal case discussed in the previous section, there are also only two independent solutions in a relaxed-MHD region containing a singular surface.
Using these results to continue across singular surfaces, relaxed-MHD basis solutions are then defined so the boundary conditions
\begin{eqnarray}
	 \zeta_{i,1,B}(r_{i-1})=1&&\zeta_{i,1,B}(r_{i})=0\nonumber
	\\
	 \zeta_{i,2,B}(r_{i-1})=0&&\zeta_{i,2,B}(r_{i})=1,\label{eq:basisboundaryEL2}
\end{eqnarray}
are satisfied. The analysis then proceeds as if there were no singular points in this region, with
$\zeta_{i,1,B}(r)$ and $\zeta_{i,2,B}(r)$ taking the place of
$\zeta_{i,1}(r)$ and $\zeta_{i,2}(r)$ in forming the matrix $A$.

\section{Two-Interface Plasma}\label{sec:twoint}

This section describes the numerical techniques used to apply the analysis to a two-interface plasma. In the plasmas considered, the
central plasma region is assumed to be a force-free Beltrami equilibrium state while the edge region between the internal interface and the plasma-vacuum interface is a pedestal region which can support a finite pressure gradient (provided we use ideal MHD in this region).
To describe the equilibrium in the edge region, the rotational transform and pressure profile were specified and the pressure variation accounted for through variation in the magnitude of the magnetic field.  For such configurations, the magnetic field takes the form
\begin{eqnarray}&\Pl_1:&\Bth(r)=c_1\J_1(\alpha_1 r)\nonumber
\\ &&\Bz(r)=c_1\J_0(\alpha_1 r)\label{eq:fieldbelpl1gen}
\\&\Pl_2:&\Bth(r)=\ibar(r)r\Bz(r)\nonumber
\label{eq:fieldpl2gen}
\\&\V:&\Bth(r)=\Bth^V/r\nonumber
\\&&\Bz(r)=\Bz^V,\label{eq:fieldvacgen}\end{eqnarray}
where $c_1$, the field strength coefficients, $\Bth^V$ and $\Bz^V$, and the Lagrange multiplier, $\alpha_1$, are constants.

Several factors do not influence stability.  First, rescaling the magnetic field by a factor, $a$, does not change the ODE for $\xir(r)$ [equation (\ref{eq:NewcombEL})] providing that the pressure profile is also rescaled by a factor of $a^2$.  As such, $B^V=\sqrt{(\Bth^V)^2+(\Bz^V)^2}$ may be set to a convenient value, in this case $\sqrt{\mu_0}$.  Second, changing the sign of both $\alpha_1$ and $\ibar(r)$ changes the sign of $\Bth(r)$ everywhere.  Since equation (\ref{eq:NewcombEL}) is unaffected by changing the sign of $k$ and $\Bth(r)$ together, only $\ibar(r_1)>0$ need be investigated, with $k$ spanning both positive and negative values.  Finally, the value of $R$ plays a limited role in stability calculations, as it determines which values of $k=-n/R$, $n$ an integer, should be considered and elsewhere appears only in the resonance condition, equation (\ref{eq:rsing}).  Without loss of generality, we select $R=1$ and make $k$ a continuous variable.   Our stability tests will therefore check if a configuration with a particular $\ibar/R$ profile is stable for all values of $R$.  If the rotational transform profile is continuous, the equilibrium is defined by the rotational transform profile in the second plasma region and the pressure profile, which may be used to determine the value of $\alpha_1$ and the vacuum field components.

It was proved by Newcomb \citea{ref:Newcomb60} that to demonstrate a configuration is stable, it is sufficient to show that it is stable for all values of $k$ when $m=1$, and for $m=0$, $k\to0$.  We have set $m$ to 1, and scanned $k$ over the range $k_{\rm min}$ to $k_{\rm max}$ in steps of $0.1$.  As it was observed that minimum values of $\lambda$ typically occur near $k=0$, $k=-\ibar_1$ or $k=-\ibar_2$, the values $k_{\rm min}=-20$ and $k_{\rm max}=20$ were chosen.  If the configuration was found to be stable, a finer scan, with a step size of $0.01$ was then performed around the least stable $k$ values.  We also set $m$ to 0 and scanned from $k=-0.5$ to $k=0.5$ in steps of $0.01$.

Three methods of stability analysis are of interest: Taylor-Taylor (possible only when the field in the edge region is Beltrami), Taylor-ideal and ideal-ideal, where the names refer to the type of analysis used in the core and second plasma regions respectively.  Solutions for the functions $\zeta_{i}$ were found using different methods for the two regions $i=1$ and $i=2$.

In the inner plasma region, the analytic solutions for $\Bb_{m,k}$ and equation (\ref{eq:xirbrrel}) provided an exact expression.  The two arbitrary constants were chosen so the function met the appropriate boundary conditions: $\zeta_{1}$ is finite at $r=0$ (ideal analysis with no singular points and Taylor analysis) or approaches $r_s$ for the small solution (ideal analysis with singular point at $r_s$), and $\zeta_{1}(r_1)=1$.

In the second plasma region, Mathematica\citea{ref:mathematica} was used to find the solutions numerically.  The methods used were different depending on the analysis used, and whether there were singular points in this region:
\begin{description}
\item{\textbf{(i) Ideal analysis with no singular points:}} the numerical solution, $\zeta^N_{2,1}(r)$ to equation (\ref{eq:NewcombEL}) that satisfies the boundary condition
\begin{eqnarray}\zeta^N_{2,1}(r_{2})=0,&&\zeta^N_{2,1}\phantom{|}\der{}(r_2)=1,
\label{eq:basisboundaryELnum}\end{eqnarray}
was found.  The value of $\zeta_{2,1}^N\phantom{|}\der{}(r_2)$ is arbitrary, and was set to unity for convenience.  $\zeta_{2,1}(r)$ is defined by
\begin{equation} \zeta_{2,1}(r)=\frac{\zeta^N_{2,1}(r)}{\zeta^N_{2,1}(r_1)}.\end{equation}  Using a similar method, $\zeta_{2,2}(r)$ was also determined.

\item{\textbf{(ii) Ideal analysis with singular point at $r_s$:}}  having determined the location of the singular point numerically, solutions either side of the singular point must be found.  For $r<r_s$, the solution $\zeta^N_{2,1,S}(r)$ to equation (\ref{eq:NewcombEL}) that satisfies the boundary conditions
\begin{eqnarray}\zeta^N_{2,1,S}(r_{s}-\epsilon)=\epsilon^{\nsing_s}&&\zeta^N_{2,1,S}\phantom{|}\der{}(r_s-\epsilon)=-\nsing_s\epsilon^{\nsing_s-1},
\label{eq:basisboundaryELnumsing}\end{eqnarray}
 was found, where $\nsing_s$ is the small value for $\nsing$ given by equation (\ref{eq:approxsoln}).  The value of $\epsilon$ was chosen so that the approximation $\xi(r)\approx |r-r_s|^{\nsing_s}$ is valid in the region $r_s-\epsilon<r<r_s$.  To do this, solutions were found for a sample configuration for various values of $\epsilon$ and the values for $\xi(r_1)$ compared.  It was found that for $\epsilon<10^{-6}$ there was little difference in the values, and so $\epsilon=10^{-7}$ was used.  The basis function $\zeta_{2,1,S}(r)$ defined in section (\ref{sec:cyl}) was then determined by dividing the function by its value at $r_1$.  Using a similar method, $\zeta_{2,2,S}(r)$ was determined from the function that satisfies the boundary conditions
\begin{eqnarray}\zeta^N_{2,2,S}(r_{s}+\epsilon)=\epsilon^{\nsing_s}&&\zeta^N_{2,2,S}\phantom{|}\der{}(r_s+\epsilon)=\nsing_s\epsilon^{\nsing_s-1}.
\label{eq:basisboundaryELnumsing2}\end{eqnarray}
\item{\textbf{(iii) Taylor analysis:}} although the analytic solutions could have been used, in this work the solutions were found numerically.  The boundary conditions satisfied by $\zeta_{2,1,B}(r)$ and $\zeta_{2,2,B}(r)$ were converted to boundary conditions for the radial component of the magnetic field using equation (\ref{eq:xirbrrel}).  The equation $\curl\BB=\alpha_2\BB$ was then solved numerically with these boundary conditions.  Equation (\ref{eq:xirbrrel}) was used again to convert back to solutions for $\zeta_{2,1,B}(r)$ and $\zeta_{2,2,B}(r)$.
\end{description}

In all cases, the solutions for $\zeta_1(r)$, $\zeta_2(r)$ and $\zeta_V(r)$ were then used to calculate the matrix elements of $A$. The eigenvalues were found, and plotted as a function of $k$.

\insertfigure{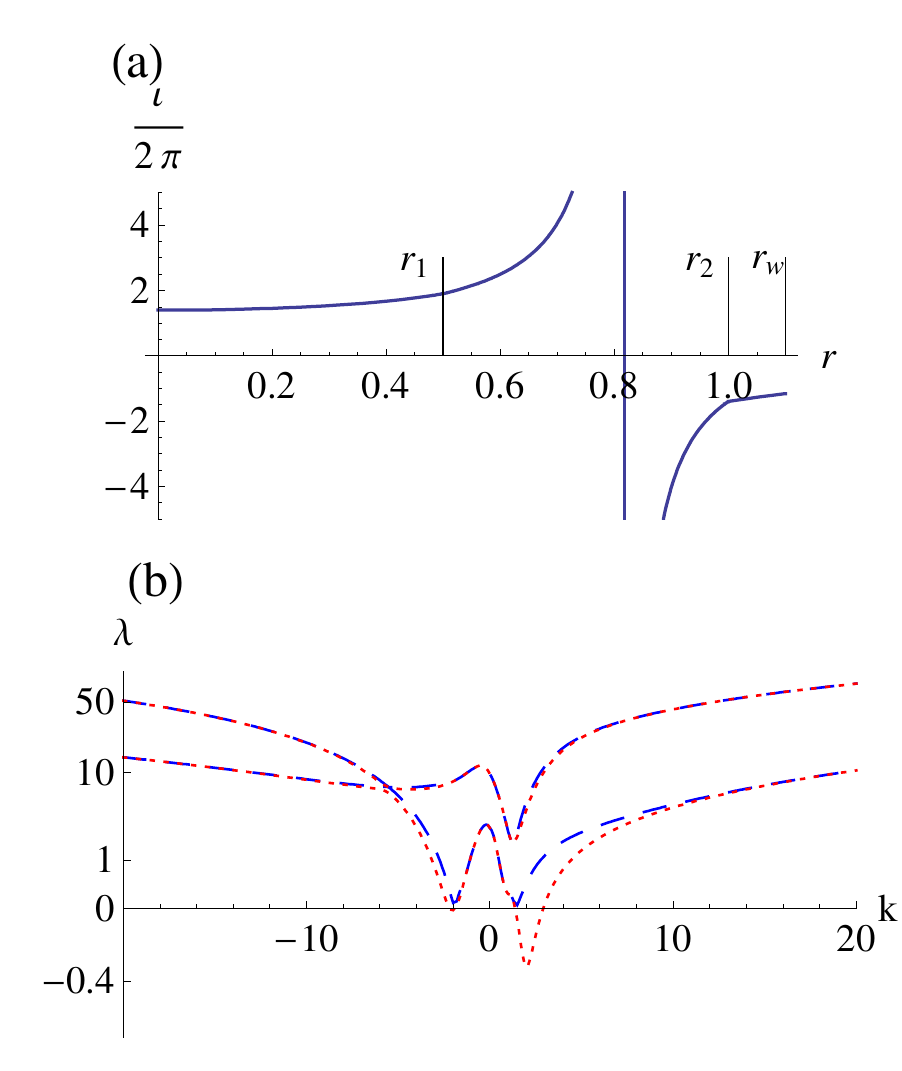}{Stability comparison of Taylor-Taylor and ideal-ideal analysis for $m=1$ modes.  Panel (a) shows the rotational transform profile used and panel (b) a dispersion curve for the Taylor-Taylor (dotted) and ideal-ideal (dashed) analysis.}  {valid}

To validate the analysis, configurations with Beltrami fields in each plasma region were analysed under both the Taylor-Taylor and ideal-ideal treatments.  The Taylor-Taylor analysis was checked by comparing an eigenvalue dispersion curve with one generated by the treatment used by Hole \emph{et al.} \citea{ref:Hole07}.  As the two methods are equivalent when no singular point is present, the ideal-ideal curve must agree with the Taylor-Taylor result for these values of $k$.  Figure (\ref{fig:valid}) (a) shows the rotational transform profile of the benchmark configuration with $r_1=0.5$, $\ibar(r_1)=1.9$ and $\ibar(r_2)=-1.4$, and figure (\ref{fig:valid}) (b) shows the $m=1$ dispersion curve.   These dispersion curves show the values of the two eigenvalues, $\lambda$, of the matrix %$M$ RLD 22/8/08 - changed M to A
$A$ as a function of $k$.  If one of the eigenvalues is less than zero, the configuration is unstable.  There is a resonant point in the edge region whenever $k>1.4$ or $k<-1.9$.  As expected, figure (\ref{fig:valid}) shows that the two methods do not agree for this range of $k$, with the ideal-ideal analysis giving the more stable eigenvalues.  If there is no resonant surface in the edge region, the two methods agree, with the exception of a small deviation, not visible on this scale, corresponding to the presence of a singular point in the core region.

\section{Limit of Vanishing Pedestal Width}\label{sec:vanint}
To resolve the vanishing interface-separation paradox, the single-interface dispersion curve was compared to the dispersion curves produced by the Taylor-Taylor and ideal-ideal treatments of the two-interface plasma in which the width of the edge region, $\Delta r=r_2-r_1$, is small.  As observed by Hole \emph{et al.} \citea{ref:Hole07} when the perfectly conducting interface is replaced by a small Taylor-relaxed region (the Taylor-Taylor treatment), the two-interface plasma is unstable where the single-interface plasma is not.  It will be demonstrated here, that when the edge region is assumed to be ideal, the one and two-interface results are in agreement.

\insertfigure{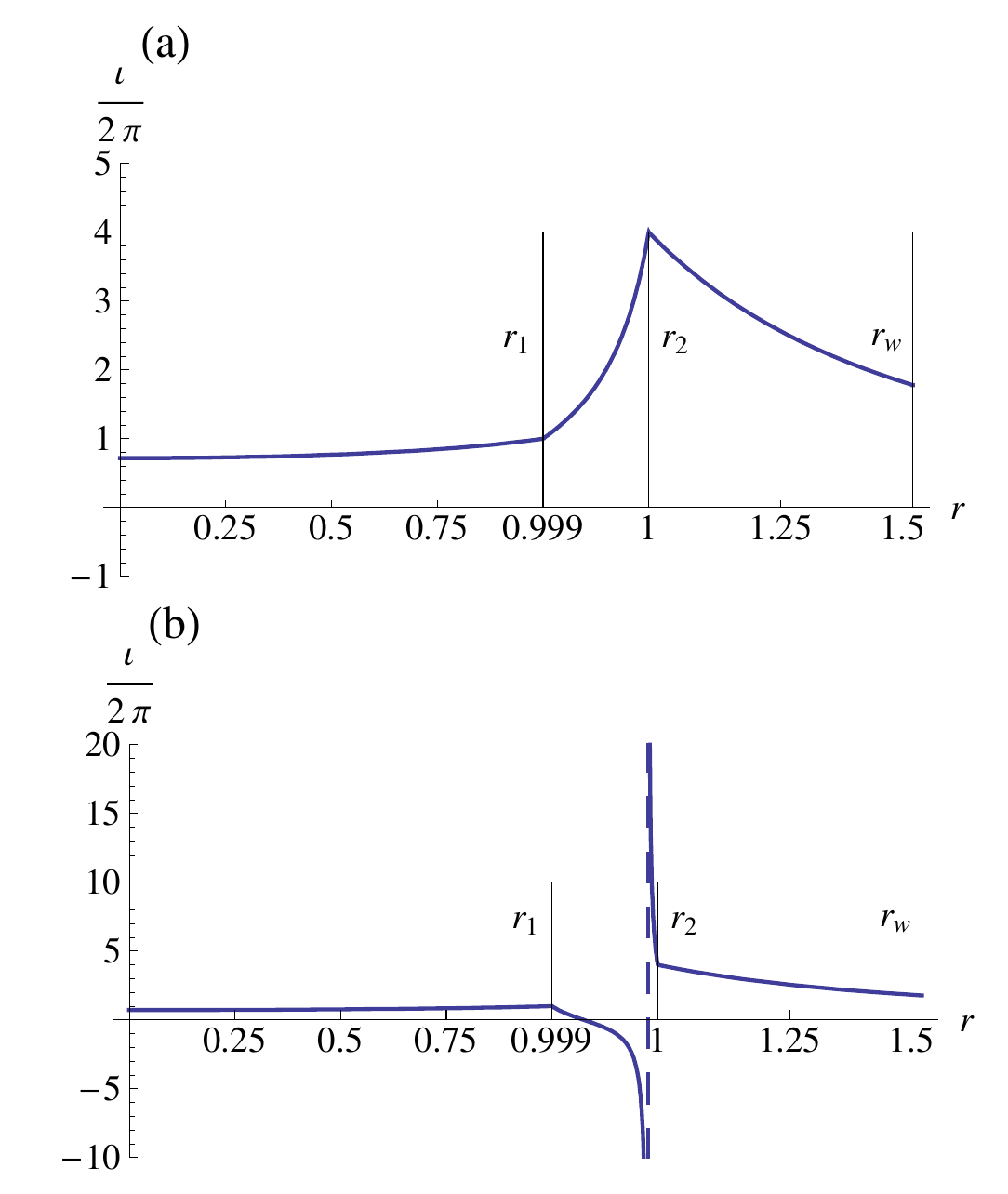}{Rotational transform profile of reference configurations used for vanishing interface separation studies.  Note the the $r$ axis has been distorted to magnify the second plasma region.  Panel (a) shows a configuration for which the rotational transform passes directly from $\ibar(r_1)=1$ to $\ibar(r_2)=4$, and panel (b) shows a configuration for which the rotational transform passed through infinity.}{refconfig}

We base our reference example on a plasma found stable by the analysis of Kaiser and Uecker: a single interface plasma with a Beltrami field, $r_w=1.5$ and rotational transform that jumps from 1 to 4 at the plasma-vacuum interface (in the parameters of Kaiser and Uecker, $\mu=1.435$ and $\delta=0.540$).\citea{ref:Kaiser04}  When the interface is replaced by a finite edge region, there are two possibilities for the rotational transform profile in the edge region.  Either it passes from $\ibar(r_1)=1$ to $\ibar(r_2)=4$ directly (hereafter referred to as the fundamental), as shown in figure (\ref{fig:refconfig}) (a), or it diverges to infinity (hereafter referred to as the harmonic), as shown on panel (b).  In the example used, the field in the edge region is Beltrami, but the conclusions hold for any rotational transform profile with these properties.  It will be shown that in each case it is mathematically possible to create a configuration with finite edge width that is stable, however such configurations are physically unrealistic.

\insertfigure{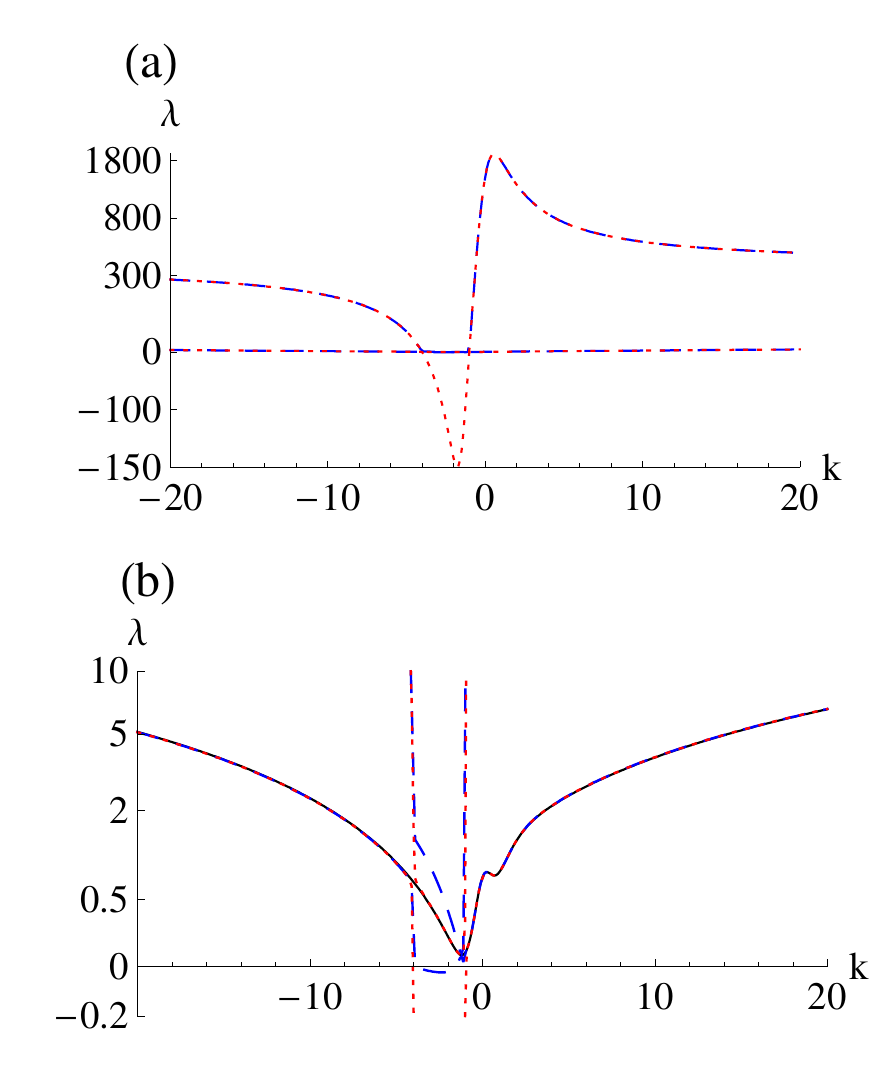}{Eigenvalue dispersion curve for $m=1$ modes of configuration with rotational transform profile show in figure (\ref{fig:refconfig}) (a), $r_1=0.999$, $r_w=1.5$ and $P(r)=0$.  Panel (a) shows the Taylor-Taylor (dotted) and ideal-ideal (dashed) eigenvalues.  Panel (b) shows the same eigenvalues, focussing on the region around marginal stability.  Also shown is half the single-interface eigenvalue (solid).}{fundstab}

\insertfigure{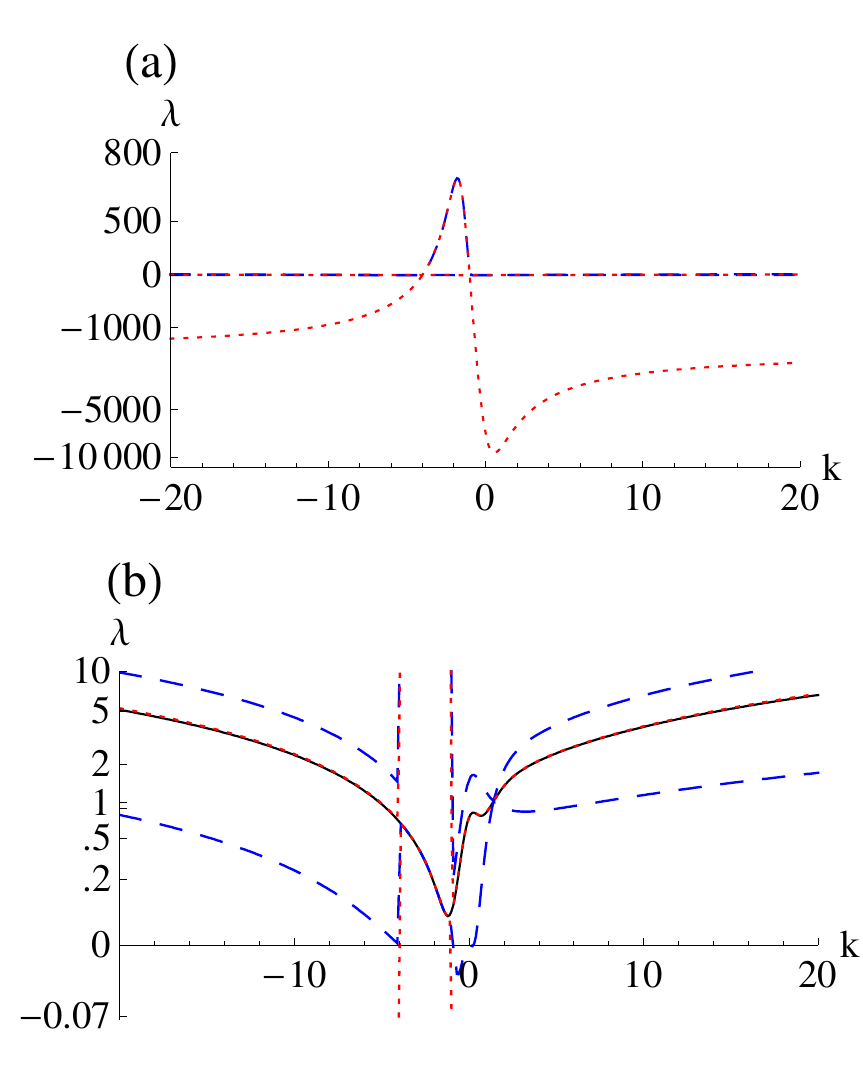}{Eigenvalue dispersion curve for $m=1$ modes of configuration with rotational transform profile show in figure (\ref{fig:refconfig}) (b), $r_1=0.999$, $r_w=1.5$ and $P(r)=0$.  Panel (a) shows the Taylor-Taylor (dotted) and ideal-ideal (dashed) eigenvalues.  Panel (b) shows the same eigenvalues, focussing on the region around marginal stability.  Also shown is half the single-interface eigenvalue (solid).}{harmstab}

Our initial stability analysis of plasmas with these rotational transform profiles has a pressure profile that is zero everywhere.  The eigenvalue dispersion curves produced using both the Taylor-Taylor and the ideal-ideal treatments for the $m=1$ mode are shown in figures (\ref{fig:fundstab}), for the fundamental, and (\ref{fig:harmstab}), for the harmonic.  Since the $m=0$ mode was stable in the limit $k\to0$, the $m=1$ dispersion curves provide a sufficient condition for stability.  In each figure, the two panels show the same dispersion curves on different scales.  Panel (b) focusses on the region around marginal stability, and includes for comparison half the single-interface eigenvalue for a configuration in which $\ibar$ jumps from $\ibar=1.9$ (inner side of interface) to $\ibar=-1.4$ (outer side of interface).  The factor of 2 arises from the normalisation, equation (\ref{eq:norm}), since the two-interface configuration contains twice the number of interfaces.  The expression for the single-interface eigenvalue, $l$, found by rearranging equation (\ref{eq:surfacexir}) using equation (\ref{eq:equilinterface}), is
\begin{equation}l=\frac{f_1(r_1)\zeta\der{1}(r_1)-f_V(r_1)\zeta\der{V}(r_1)}{r_1} -\frac{2k^2 r_1\Delta P}{m^2+k^2r_1^2},\label{eq:eigN1}\end{equation}
 where $r_1$ is the location of the interface and $\Delta P$ is the pressure on the inner side of the interface minus the pressure on the outer side.

Looking first at the fundamental, under the Taylor-Taylor analysis, one of the eigenvalues agrees with half the single-interface result, while the second eigenvalue is very positive for $k<-4$ and $k>-1$ and very negative for $-4<k<-1$, the latter result being responsible for the conclusion of instability.  The range $-4<k<-1$ also corresponds to the values of $k$ at which there is a resonant surface in the edge region.  Outside this range, the two methods produce the same eigenvalues, as is expected.  Within this range, neither ideal-ideal eigenvalue agrees with the single-interface case, and one eigenvalue is negative for $-4<k<-1.5$, meaning the configuration is unstable.

For the harmonic, we find that one Taylor-Taylor eigenvalue agrees with half the single interface result, while the other is very positive for $-4<k<-1$ and very negative for $k<-4$ and $k>-1$.  The range of $k$ values for which the Taylor-Taylor analysis is negative again corresponds to the $k$ values at which there is a resonant surface in the edge region.  The ideal-ideal analysis has negative eigenvalues for $-1<k<0$.

To understand why the ideal-ideal analysis may produce negative eigenvalues when the single-interface result is stable, the expressions for the the eigenvalues, at $k$ values with resonant surface in the edge region in the limit $\Delta r\to0$,
 \begin{equation}\lambda_1=\frac{f_1(r_1)\zeta\der{1}(r_1)}{r_1}+\frac{2k^2r_2^2}{k^2r_2^2+m^2}[P(r_s)-P(r_1)],\label{eq:lambda1app}\end{equation}
\begin{equation}\lambda_2=\frac{2k^2r_2^2}{k^2r_2^2+m^2}[P(r_2)-P(r_s)]-\frac{f_V(r_2)\zeta\der{V}(r_2)}{r_2},\label{eq:lambda2app}\end{equation}
will be examined.  The expressions are equations (93a) and (93b) of Newcomb \citea{ref:Newcomb60}, converted to the notation used here, and multiplied by $1/(k^2r_2^2+m^2)$.  If the configuration is stable, both eigenvalues must be positive, and the expression may be rearranged to yield
\begin{equation}-\frac{f_V(r_2)\zeta\der{V}(r_2)}{r_2}\frac{k^2r_2^2+m^2}{2k^2r_2^2}>P(r_s(k))>P(r_1)-\frac{f_1(r_1)\zeta\der{1}(r_1)}{r_1}\frac{k^2r_2^2+m^2}{2k^2r_2^2},\label{eq:P(k)cond}\end{equation}
where it was noted that $P(r_2)=0$.  The pressure at the singular points is written in the form $P(r_s(k))$ to emphasise that the pressure bounds are determined not by the position of the singular point, $r_s$, but by the value of $k$ with which it is resonant.  Put another way, if we know that the rotational transform varies between the values $\ibar_1$ and $\ibar_2$, and which harmonic it follows, equation (\ref{eq:P(k)cond}) gives information about the form a stable pressure profile must take, without it being necessary to know the precise location of any of the singular points.

\insertfigure{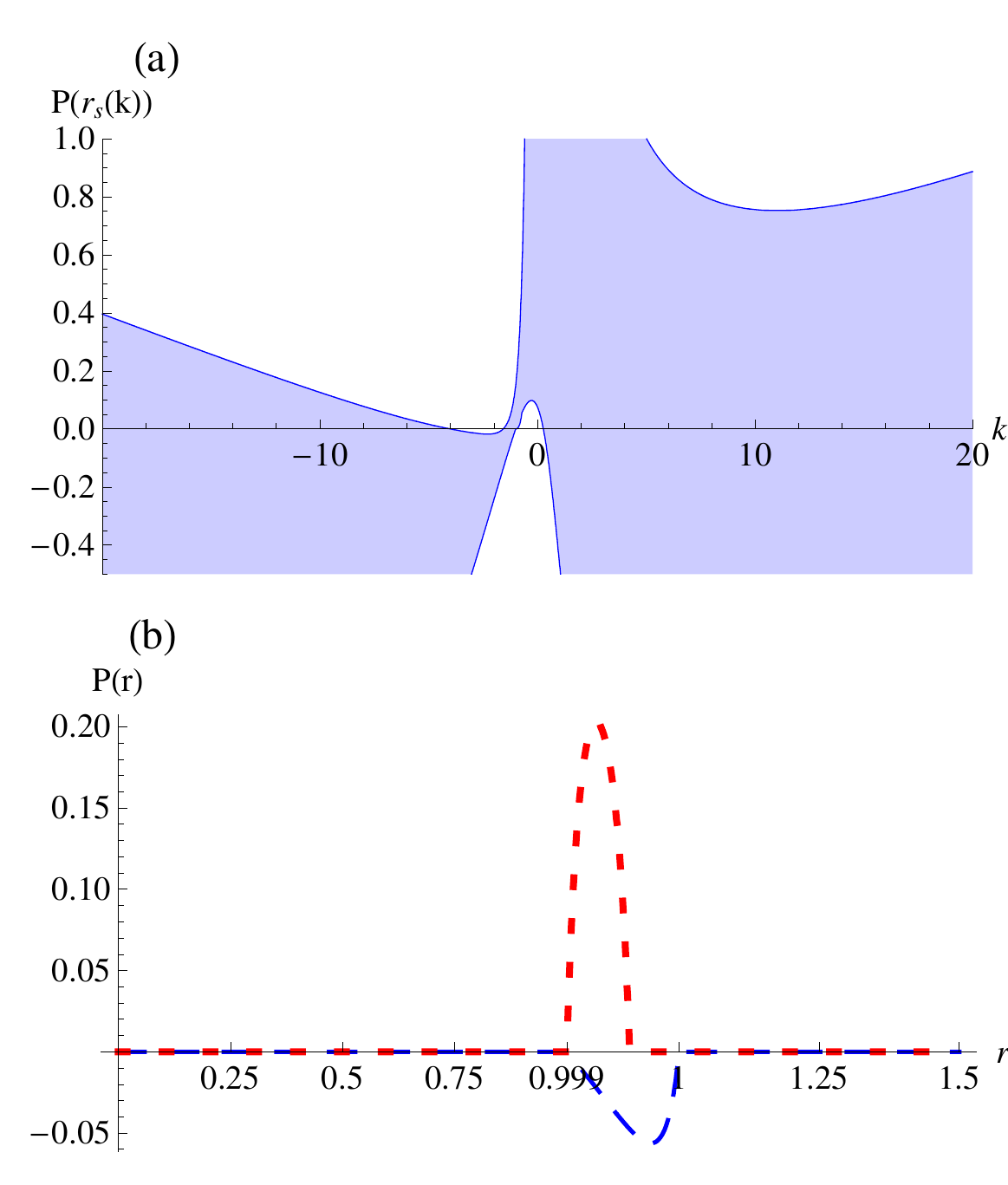}{(a) Stable region (shaded) of $P(r_s(k))$ under ideal-ideal analysis for configurations with $\ibar_1=1$, $\ibar_2=4$, $r_0=0.999$ and $r_w=1.5$.  (b) Stabilising pressure profile for fundamental, shown in figure (\ref{fig:refconfig}) (a), (dashed) and harmonic, shown in figure (\ref{fig:refconfig}) (b), (dotted) configurations.  Note that the $r$ axis has been distorted to magnify the second plasma region.  }{pressure}

To illustrate the significance of these requirements, figure (\ref{fig:pressure}) (a)
shows the lower and upper bounds of equation (\ref{eq:P(k)cond}) as a function of $k$ when $\ibar_1=1$, $\ibar_2=4$ and $P(r_1)=0$, the configuration used to produce the plots of figure (\ref{fig:refconfig}).  This figure reveals that if the rotational transform is resonant with $k$ values in the range $-4<k<-1.5$, the plasma may only be stable if the pressure is negative at the radial position at which these resonances occur.  As the fundamental must be resonant with this range of $k$ values in the edge region, this class of configurations may only be stabilised by allowing negative pressure.  Since the configuration analysed in figure (\ref{fig:fundstab}) had zero pressure everywhere, this explains why the configuration was unstable over this range of $k$.  Similarly, the pressure must be positive wherever resonances with $k$ values in the range $-1<k<.5$ occur.  Since harmonic configurations must have such resonances, the harmonic may only be stable if the pressure in the edge region reaches a greater value than that on the interface.

When the stability analysis was repeated for the pressure profiles shown in figure (\ref{fig:refconfig}), chosen as example profiles that satisfy the pressure bound, the configurations were found to be stable for $m=1$ modes and $m=0$ in the limit $k\to0$.  It is therefore mathematically possible to construct a pressure profile that stabilizes the configuration with $\ibar_1=1$, $\ibar_2=4$ and $r_w=1.5$, in agreement with the conclusion of stability reached by Kaiser and Uecker.  However, to achieve this it was necessary to either allow the plasma to have negative pressure, or a very narrow pressure maximum at the edge, both of which are physically unlikely.

Rearranging inequality (\ref{eq:P(k)cond}) reproduces the inequality $l>0$, the single-interface stability condition of equation (\ref{eq:eigN1}).  Therefore, under the ideal-MHD treatment, the stability of a two-interface plasma in the limit $\Delta r\to0$ may be reconciled with the stability of the corresponding single-interface plasma.  Specifically, if the single-interface eigenvalue is greater than zero for all values of $k$, then: (i) if there is no resonant surface between the two interfaces, one of the eigenvalues will agree with half the single-interface eigenvalue, while the other is extremely positive, and (ii) if there is a singular point between the two surfaces, it is possible to construct a pressure profile using inequality (\ref{eq:P(k)cond}) for which both eigenvalues are positive.  If the single-interface configuration is unstable, then the two-interface configuration will be unstable under both Taylor-Taylor and ideal-ideal treatments.

Although we have shown that it is mathematically possible to reconcile the one and two interface stability results, the requirements placed on the pressure of the two interface plasma may be physically implausible.  It is therefore not sufficient to simply apply the single interface analysis of Kaiser and Uecker.  It is also not necessary to have a detailed knowledge of the plasma in the edge region.  Rather, one must consider the field and pressure profiles of the core and vacuum regions and the resonances in the edge region, and determine if the restrictions they place on stable pressure profiles are physically acceptable.

Although it was not possible to construct a physically reasonable configuration from the single interface plasma considered here, this is not true for all single interface plasmas considered by Kaiser and Uecker.  Stable reverse field pinch configurations, with rotational transform profile similar to that of figure (\ref{fig:valid}) (a), but with a vanishing edge region, may be constructed with pressure that is zero everywhere.

\section{Conclusion}We have developed a unified energy principle for analysing the MHD stability of plasmas consisting of both Taylor-relaxed and ideal MHD regions. The gauge $\Ba = \Bxi\cross\BB$ for the vector potential, $\Ba$, of linearized perturbations is used, with the equilibrium magnetic field $\BB$ obeying a Beltrami equation, $\curl\BB = \mu\BB$, in relaxed regions.  A central result is that $\Bxi$ obeys the \emph{same} Euler-Lagrange equation whether ideal or relaxed MHD is used for perturbations, except in the neighbourhood of the magnetic surfaces where $\BB\dotv\grad$ is singular.  To explore this difference, we developed a procedure for constructing global multi-region solutions in cylindrical geometry.  Each region is a cylindrical shell in which the plasma is either ideal MHD or Taylor-relaxed.  The shells are nested and separated by ideal MHD barriers or interfaces of zero width.  In this configuration, only Newcomb's small solutions are allowed for ideal MHD regions whereas in relaxed MHD only the odd-parity large solution and even-parity small solution are allowed.

By way of illustration, we have evaluated stability implications using the different energy principles to a two interface plasma whose rotational transform profile resembles that of a reverse field pinch. A second motivation to study this configuration is the resolution of singular-limit problem encountered previously by Hole \emph{et al} \citea{ref:Hole07} in multi-region relaxed MHD,when two barrier interfaces become arbitrarily close. In that work, it was found that the stability of a two interface plasma with Taylor-relaxed regions does not approach the stability of a single interface configuration in the limit that the interface separation vanishes.

We infer that the two interface configuration is unstable to a tearing mode, and that if the Taylor-relaxed region is replaced by an ideal MHD region of nonzero width, stability conclusions of the single interface and vanishing barrier region agree.  By solving for the eigenvalues in the ideal treatment in the limit of vanishing barrier width, and comparing these to zero, we have put brackets on the pressure at field resonances.  Using this, we have shown that providing the magnetic shear is monotonic across the plasma, it is always mathematically possible to construct a stable non-pathological pressure profile to connect a pressure jump.  If the shear toggles sign however, such that an ideal resonance is in the plasma, stabilisation in the barrier region may require a pathological pressure profile (ie negative pressure).

A conclusion of our work is that the stability of plasmas with jump in rotational transform across a barrier can not be concluded from a barrier treatment (such as Kaiser and Uecker or Hole et al) alone.  One must also identify whether resonances exist in the barrier in the limit it is spatially resolved.  If no resonances exist in the barrier and the plasma is stable, the structure of the pressure jump across the barrier in the limit it is resolved is irrelevant. If resonances exist in the barrier stability is then a function of the pressure profile within the barrier in the limit it is spatially resolved.

\bibliography{referenceexample2}
\bibliographystyle{vancouver}

\end{document}